\magnification=1200
\hfuzz=3pt
\hsize=12.5cm
\hoffset=0.32cm
\baselineskip=18pt
\voffset=\baselineskip
\nopagenumbers
\font\title=cmssdc10 at 20pt
\font\smalltitle=cmssdc10 at 14pt

\font\tenbm=cmmib10

\font\sevenbm=cmmib7

\font\fivebm=cmmib5
\newfam\bmfam
\textfont\bmfam=\tenbm
\scriptfont\bmfam=\sevenbm
\scriptscriptfont\bmfam=\fivebm
\def\bm{\fam\bmfam\tenbm}

\let\math=\mathchardef
\math\Gamma="7000 
\math\Delta="7001 
\math\Theta="7002 
\math\Lambda="7003 
\math\Xi="7004 
\math\Pi="7005 
\math\Sigma="7006  
\math\Upsilon="7007 

\math\Phi="7008 
\math\Psi="7009 
\math\Omega="700A  
\math\alpha="710B 
\math\beta="710C 
\math\gamma="710D 
\math\delta="710E 
\math\epsilon="710F 

\math\zeta="7110 
\math\eta="7111 
\math\theta="7112 
\math\iota="7113 
\math\kappa="7114 
\math\lambda="7115 
\math\mu="7116 
\math\nu="7117 

\math\xi="7118 
\math\pi="7119 
\math\rho="711A 
\math\sigma="711B 
\math\tau="711C 
\math\upsilon="711D 
\math\phi="711E 
\math\chi="711F 

\math\psi="7120 
\math\omega="7121 
\math\varepsilon="7122 
\math\vartheta="7123 
\math\varpi="7124 
\math\varrho="7125 
\math\varsigma="7126 
\math\varphi="7127

\let\text=\textstyle
\let\display=\displaystyle

\def\IM{\mathop{\Im m}\nolimits}
\def\sectionstyle{\smalltitle}
\newskip\beforesectionskip
\newskip\aftersectionskip
\beforesectionskip=4mm plus 1mm minus 1mm
\aftersectionskip=2mm plus .2mm minus .2mm
\newcount\mysectioncounter
\def\resetsections{\mysectioncounter=0}
\resetsections
\newcount\myeqcounter
\def\mysection#1\par{\par\removelastskip\penalty -250
\vskip\beforesectionskip
\global\advance\mysectioncounter by 1\noindent
\myeqcounter=0{\sectionstyle\the\mysectioncounter.
#1}\par
\nobreak\vskip\aftersectionskip}
\def\myeqno{\global\advance\myeqcounter by 1\eqno{(\the\mysectioncounter.\the\myeqcounter)}}
\def\mydispeqno{\global\advance\myeqcounter by 1\hfill\llap{(\the\mysectioncounter
.\the\myeqcounter)}}
\headline={\hfil\tenrm\folio\hfil}
\footline={\hfil}
\baselineskip=12pt
\parindent= 20pt
\centerline{\smalltitle Quantum fluctuation-dissipation theorem: a time domain formulation}
\vskip 0.5cm
\centerline{No\"elle POTTIER}  
\centerline{\sl Groupe de Physique des Solides\footnote{$^1$}{\rm Laboratoire
associ\'e au C.N.R.S. (U.M.R. n$^0\ 7588$) et aux Universit\'es Paris 7 et Paris
6.}, Universit\'e Paris 7,}
\centerline{\sl Tour 23, 2 place Jussieu, 75251 Paris Cedex 05, France,\/}
\smallskip
\centerline{and}
\smallskip
\centerline{Alain MAUGER}
\centerline{\sl Laboratoire des Milieux D\'esordonn\'es et
H\'et\'erog\`enes\footnote{$^2$}{\rm Laboratoire associ\'e au C.N.R.S. (U.M.R. n$^0\ 7603$) et \`a
l'Universit\'e Paris 6.},}
\centerline{\sl Universit\'e Paris 6, Tour 22, Case 86,}
\centerline{\sl 4 place Jussieu, 75252 Paris Cedex 05, France.\/}
\vskip 2cm
\noindent
{\smalltitle Abstract}
\medskip
\noindent
A time-domain formulation of the equilibrium quantum fluctuation-dissipation theorem (FDT) in the
whole range of temperatures is presented. In the classical limit, the FDT establishes a
proportionality relation between the dissipative part of the linear response function and the
derivative of the corresponding equilibrium correlation function. At zero temperature,
the FDT takes the form of Hilbert transform relations between the dissipative part of the response
function and the corresponding symmetrized equilibrium correlation function, which allows to
establish a connection with analytic signal theory. The time-domain formulation of the FDT is
especially valuable when out-of-equilibrium dynamics is concerned, as it is for instance the case
in the discussion of aging phenomena. 

\vskip 1cm
\parindent=0pt
{\bf PACS numbers:} 

05.30.-d Quantum statistical mechanics.

\medskip
{\bf KEYWORDS:} 
\medskip
{\sl Corresponding author:

No\"elle POTTIER
  
Groupe de Physique des Solides, Universit\'e Paris 7,

Tour 23, 2 place Jussieu, 75251 Paris Cedex 05, France

Fax number: +33 1 43 54 28 78. E-mail: pottier@gps.jussieu.fr\/}
\vfill
\break
\parindent=20pt
\baselineskip=12pt

\mysection{Introduction}

The fluctuation-dissipation theorem (FDT), valid for dynamic variables in equilibrium, is usually
written in a form which involves generalized susceptibilities and spectral densities, which are
frequency-dependent quantities [1]-[4]. However, in order to discuss certain time-dependent
properties, it may be more convenient to have at hand a formulation of the theorem in the time
domain. This need is for instance well illustrated in the discussion of aging effects in
response and/or correlation functions of out-of-equilibrium dynamic variables. Then, the
equilibrium FDT is {\sl a priori\/} not applicable and has to be modified by the introduction
of a violation factor or of an effective temperature, which can conveniently be defined in
terms of time-dependent quantities [5]-[8].

Following arguments outlined in [3], we develop below in a direct and simple manner different
ways in which the equilibrium quantum FDT can be formulated in the time domain. The corresponding
expressions of the theorem, which are established in the whole range of temperatures, allow in
particular for a discussion of both the classical limit [3] and the zero-temperature case. In this
latter situation, an interesting connection, which does not seem to have been put forward
previously, is shown to exist with analytic signal theory [9]-[10]. 

\mysection{Time domain formulation of the equilibrium FDT}

Let us consider here a system in equilibrium described in the absence of external perturbations by
a time-independent hamiltonian $H_0$. In the following we will be concerned with equilibrium
average values which we will denote as $\langle\ldots\rangle$, the symbol
$\langle\ldots\rangle$ standing for ${\rm Tr}\,\rho_0\ldots$, with $\rho_0={e^{-\beta H_0}/{\rm
Tr}\,e^{-\beta H_0}}$. 

Since we intend to discuss about linear response functions and symmetrized equilibrium  
correlation functions generically denoted as $\tilde\chi_{BA}(t,t')$ and $\tilde
C_{BA}(t,t')$,  we shall assume that the observables of interest $A$ and $B$ do not commute with
$H_0$ (were it the case, the response function $\tilde\chi_{BA}(t,t')$ would indeed be zero). This
hypothesis implies in particular that $A$ and $B$ are centered : $\langle A\rangle=0$,
$\langle B\rangle=0$.

Generally speaking, the time domain formulations of the equilibrium FDT establish the link between
the linear response function $\tilde\chi_{BA}(t,t')$ and the symmetrized equilibrium  correlation
function $\tilde C_{BA}(t,t')={1\over 2}[\langle A(t')B(t)\rangle+\langle B(t)A(t')\rangle]$ (or the
derivative ${\partial\tilde C_{BA}(t,t')/\partial t'}$).
\bigskip
{\bf 2.1. The response function in terms of the correlation function} 

Consider two quantum-mechanical observables $A$ and $B$ with thermal equilibrium correlation
functions verifying the property
$$\langle A(t'-i\hbar\beta)B(t)\rangle=\langle B(t)A(t')\rangle,\qquad\beta={1/kT},\myeqno$$
and compute the contour integral 
$$I=\oint_{\Gamma}\langle
A(\tau)B(t)\rangle\,{\pi\over\beta\hbar}\,{1\over\sinh{\pi(\tau
-t')\over\beta\hbar}}\,d\tau\myeqno$$ where $t$ is a real time and $\Gamma$ the closed contour in
the complex $\tau$-plane represented on Fig.~1. One checks easily that the integrand in $I$ does
not present any singularities inside
$\Gamma$. Actually, denoting by $|\lambda\rangle$ and $E_\lambda$ the eigenstates and eigenenergies
of the system hamiltonian $H_0$, one has
$$\langle A(\tau)B(t)\rangle={1\over
Z}\sum_{\lambda,\lambda'}A_{\lambda\lambda'}B_{\lambda'\lambda}\,e^{-\beta
E_\lambda+i(\tau-t)(E_\lambda-E_{\lambda'})},\qquad Z={\rm Tr}\,e^{-\beta H_0},\myeqno$$
which displays the fact that the continuation to the region $-\beta\leq\IM\tau\leq 0$
of the complex $\tau$-plane of the correlation function $\langle A(\tau)B(t)\rangle$ is
analytic [3], [11].

According to the Cauchy theorem, the integral $I$ is thus equal to zero. In the limit
$R\to\infty$, one obtains, by gathering the various contributions to $I$, the relation
$$\displaylines{i\pi[\langle B(t)A(t')\rangle-\langle A(t')B(t)\rangle]=\hfill\cr
\hfill{\rm
vp}\int_{-\infty}^\infty dt''\,[\langle
A(t'')B(t)\rangle+\langle
B(t)A(t'')\rangle]\,{\pi\over\beta\hbar}\,{1\over\sinh{\pi(t''-t')\over\beta\hbar}},
\mydispeqno\cr}$$
where the symbol ${\rm vp}$ denotes the Cauchy principal value. Taking into account the Kubo formula
for the response function, one gets from Eq. (2.4) the expression
$$\tilde\chi_{BA}(t,t')={2\over\pi\hbar}\,\Theta(t-t')\,{\rm vp}\int_{-\infty}^\infty dt''\,
\tilde C_{BA}(t,t'')\,{\pi\over\beta\hbar}\,{1\over\sinh{\pi(t''-t')\over\beta\hbar}},\myeqno$$
where $\Theta(t)$ denotes the unit step-function.

Eq. (2.5) allows to compute the response function $\tilde\chi_{BA}$ in terms of the symmetrized
correlation function $\tilde C_{BA}$. Then, introducing the dissipative part $\tilde\xi_{BA}$ of
$\tilde\chi_{BA}$, as defined by
$$\tilde\chi_{BA}(t,t')=2i\,\Theta(t-t')\,\tilde\xi_{BA}(t,t'),\myeqno$$
one gets from Eq. (2.5):
$$i\hbar\,\tilde\xi_{BA}(t,t')={1\over\pi}\,{\rm vp}\int_{-\infty}^\infty dt''\,
\tilde C_{BA}(t,t'')\,{\pi\over\beta\hbar}\,{1\over\sinh{\pi(t''-t')\over\beta\hbar}}.\myeqno$$
Since the two-time equilibrium averages involved in $\tilde\xi_{BA}$ and $\tilde C_{BA}$ only depend
on the time differences involved, Eq. (2.7) can be rewritten as a convolution product, that is
$$i\hbar\,\tilde\xi_{BA}(t)={1\over\pi}\,\tilde C_{BA}(t)*{\rm
vp}\,{\pi\over\beta\hbar}\,{1\over\sinh{\pi t\over
\beta\hbar}},\myeqno$$
the convolution in the r.h.s. being taken with respect to~$t$.  
\vfill
\break
{\bf 2.2. The correlation function in terms of the response function}

The expression of the correlation function in terms of the response function can be derived, either
in the same way as above ({\sl i.e.\/} by using contour integration), or by inverting the
convolution product (2.8). 

Compute on the contour $\Gamma$ (Fig.~1) the integral
$$J=\oint_{\Gamma}\langle
A(\tau)B(t)\rangle\,{\pi\over\beta\hbar}\,\coth{\pi(\tau-t')\over\beta\hbar}\, d\tau.\myeqno$$
Using similar arguments as above ({\sl i.e.\/} noticing that
$J=0$ since there are no singularities of the integrand of
$J$ inside
$\Gamma$), one obtains the relation
$$\displaylines{{1\over 2}[\langle A(t')B(t)\rangle+\langle B(t)A(t')\rangle]=\hfill\cr
\hfill -{i\over 2\pi}\,{\rm
vp}\int_{-\infty}^\infty dt''\,[\langle
B(t)A(t'')\rangle-\langle A(t'')B(t)\rangle]\,{\pi\over\beta\hbar}\,\coth{\pi(t''-t')\over
\beta\hbar},\mydispeqno\cr}$$
that is
$$\tilde C_{BA}(t,t')=-{\hbar\over 2\pi}\,{\rm vp}\int_{-\infty}^\infty
dt''\,\bigl[\tilde\chi_{BA}(t,t'')-\tilde\chi_{AB}(t'',t)\bigr]\,{\pi\over\beta\hbar}
\,\coth{\pi(t''-t')\over\beta\hbar},\myeqno$$
or
$$\tilde C_{BA}(t,t')=-{1\over\pi}\,{\rm
vp}\,\int_{-\infty}^\infty dt''\,i\hbar\,\tilde\xi_{BA}(t,t'')\,{\pi\over\beta\hbar}
\,\coth{\pi(t''-t')\over\beta\hbar}.\myeqno$$

Before going further, let us add a comment. The arguments which have previously been used in the
computation of the integral $I$ must be refined in order to show, first, that the integrals in the
r.h.s. of Eqs. (2.10)-(2.12) are properly defined, and, second, that the contributions to the
integral $J$ (Eq. (2.9)) of the two vertical segments of abscissas $R$ and $-R$ of the contour
$\Gamma$ (Fig.~1) do vanish in the limit $R\to\infty$. The question stems from the fact that the
function $\coth({\pi t/\beta\hbar})$ tends towards a finite limit for large values of its argument,
contrary to the function ${1/\sinh}({\pi t/\beta\hbar})$ involved in the computation of $I$, which
tends towards zero (this latter property insuring that the integrals in the r.h.s. of Eqs. (2.4),
(2.5) and (2.7) are properly defined and that the  contribution to $I$ of the above-mentioned
segments is actually zero in the limit $R\to\infty$). 

A sufficient condition is $\lim_{R\to\infty}\langle A(\pm
R-i\hbar y)B(t)\rangle=\langle A\rangle\langle B\rangle$ ($=0$ since $A$ and $B$ are
centered)\footnote{$^1$}{Note that for $y=0$ this condition amounts to
$\lim_{R\to\infty}\langle A(\pm R)B(t)\rangle=0$ while for $y=\beta$ it amounts to
$\lim_{R\to\infty}\langle B(t)A(\pm R)\rangle=0$.}. Thus, when treating systems with a finite number
of degrees of freedom, and oscillating response and correlation functions, in which case this limit
does not even exist, it will be convenient to introduce a small damping which will be eventually
let equal to zero, or, which amounts to the same, to treat the response and correlation functions as
distributions. Further details will be provided when treating the harmonic oscillator example
(Section 2.4).

Eq. (2.12) can be viewed as the reciprocal of Eq. (2.7), since it allows to compute the
symmetrized correlation function $\tilde C_{BA}$ in terms of the dissipative part $\tilde\xi_{BA}$
of $\tilde\chi_{BA}$. It can be rewritten, using convolution product
notations, as
$$\tilde C_{BA}(t)=
-{1\over\pi}\,i\hbar\,\tilde\xi_{BA}(t)*{\pi\over\beta\hbar}\,{\rm
vp}\,\coth{\pi t\over\beta\hbar}.
\myeqno$$
Note that this latter expression can also be obtained directly by inverting the convolution product
(2.8), which is easily done by using the relation
$${\pi\over\beta\hbar}\,{\rm vp}{1\over\sinh{\pi t\over\beta\hbar}}*{\pi\over\beta\hbar}\,{\rm
vp}\,\coth{\pi t\over\beta\hbar}=-\pi^2\,\delta(t),\myeqno$$
demonstrated in Appendix A.

Eq. (2.8) together with the inverse relation (2.13) constitute a formulation of the equilibrium
FDT in the time domain. 
\bigskip
{\bf 2.3. Relation with the usual frequency domain formulation}
 
Eqs. (2.8) and (2.13) just correspond by Fourier transformation to the usual fluctuation-dissipation
relations between the dissipative part $\xi_{BA}(\omega)$ of the
susceptibility and the Fourier transform $C_{BA}(\omega)$ of the symmetrized correlation function
[1]-[4]:
$$\xi_{BA}(\omega)={1\over\hbar}\,\tanh{\beta\hbar\omega\over 2}\,C_{BA}(\omega),\qquad
C_{BA}(\omega)=\hbar\,\coth{\beta\hbar\omega\over 2}\,\xi_{BA}(\omega).\myeqno$$ 

Indeed, taking the Fourier transform of formulas (2.15), with the following definition of the
Fourier transformation $F(\omega)=\int_{-\infty}^\infty dt\,\tilde
F(t)\,e^{i\omega t}$,
$\tilde F(t)={1\over 2\pi}\int_{-\infty}^\infty d\omega\,C_{BA}(\omega)\,e^{-i\omega t}$, and
making use of the Fourier formulas ($A$.3) and ($A$.6), one obtains Eqs. (2.8) and (2.13),
rewritten below for clarity:
$$i\hbar\,\tilde\xi_{BA}(t)={1\over\pi}\,\tilde C_{BA}(t)*{\pi\over\beta\hbar}\,{\rm vp}\,
{1\over\sinh{\pi t\over\beta\hbar}},\ \tilde
C_{BA}(t)=-{1\over\pi}\,i\hbar\,\tilde\xi_{BA}(t)*{\pi\over\beta\hbar}\,
{\rm vp}\,\coth{\pi t\over\beta\hbar}.\myeqno$$
\bigskip
{\bf 2.4. A basic example : the harmonic oscillator}

Let us consider as a basic example a harmonic oscillator of mass $m$ and angular frequency
$\omega_0$, in thermal equilibrium at temperature $T$. The oscillator displacement being denoted by
$x$, one has
$$\langle x(t)x\rangle={\hbar\over
2m\omega_0}\,\left[(1+n)\,e^{-i\omega_0t}+n\,e^{i\omega_0t}\right]\myeqno$$
and 
$$\langle xx(t)\rangle={\hbar\over
2m\omega_0}\,\left[n\,e^{-i\omega_0t}+(1+n)\,e^{i\omega_0t}\right],\myeqno$$
where $n={1/(e^{\beta\hbar\omega_0}-1)}$ is the Bose-Einstein function at temperature
$T$.
One deduces from Eqs. (2.17) and (2.18) the expressions of $\tilde\xi_{xx}(t)$ and $\tilde
C_{xx}(t)$ :

$$i\hbar\,\tilde\xi_{xx}(t)={\hbar\over 2m\omega_0}\,\sin\omega_0t\myeqno$$
and 
$$\tilde C_{xx}(t)={\hbar\over
2m\omega_0}\,\coth{\beta\hbar\omega_0\over
2}\,\cos\omega_0t.\myeqno$$

The fluctuation-dissipation relations (2.8) and (2.13) respectively read:
$$\sin\omega_0t={1\over\pi}\,\coth{\beta\hbar\omega_0\over
2}\,\cos\omega_0t*{\pi\over\beta\hbar}\,{\rm vp}\,{1\over\sinh{\pi t\over\beta\hbar}}\myeqno$$
and
$$\coth{\beta\hbar\omega_0\over
2}\,\cos\omega_0t=-{1\over\pi}\,\sin\omega_0t*{\pi\over\beta\hbar}\,{\rm
vp}\,\coth{\pi t\over\beta\hbar}.\myeqno$$
Both formulas can be easily checked by a direct calculation, which we report in Appendix B,
together with supplementary details\footnote{$^2$}{See the remarks in Section 2.2.} on the contour
integration needed to compute the integral $J$ (Eq. (2.9)). 

Now, coming back to the general case, let us discuss, first the classical limit, then the
zero-temperature case.

\mysection{The classical limit}

Before entering the discussion of this limit, we shall propose another form of the first FDT
relation (Eqs. (2.7) or (2.8)) in the time domain. It gives the expression of the linear response
function in terms of the derivative of the equilibrium correlation function [3], and reveals to be
especially useful when studying the classical limit.
\bigskip
{\bf 3.1. Expression of the response function in terms of the derivative of the correlation
function}

Integrating by parts, one can recast Eq. (2.7) into the equivalent form
$$i\hbar\,\tilde\xi_{BA}(t,t')={1\over\pi}\int_{-\infty}^\infty dt''\,{\partial
\tilde C_{BA}(t,t'')\over\partial t''}\,\log\Bigl|\coth{\pi(t''-t')\over
2\beta\hbar}\Bigr|,\myeqno$$
or, making use of convolution product notations since $\tilde\xi_{BA}$ and $\tilde C_{BA}$ only
depend on the time differences involved,
$$i\hbar\,\tilde\xi_{BA}(t)=-{1\over\pi}\,{d\tilde
C_{BA}(t)\over dt}*\log\Bigl|\coth{\pi t\over
2\beta\hbar}\Bigr|,\myeqno$$
an expression equivalent to Eq. (2.8). Thus, at any temperature, the dissipative part of the
response function ({\sl i.e.\/} the function $\tilde\xi_{BA}(t)$) appears to be proportional to the
convolution product taken with respect to $t$ of the functions ${d\tilde C_{BA}(t)/dt}$ and
$\log|\coth({\pi t/2\beta\hbar})|$. This latter function is very peaked around $t=0$ at high
temperature while it becomes more and more spread around this value as the temperature decreases.
\bigskip
{\bf 3.2. The classical limit}

In the classical limit, making use of the property ($A$.14), that is
$$\log|\coth\,ax|\sim_{|a|\to\infty}{\pi^2\over 4|a|}\,\delta(x),\myeqno$$
with $a={\pi/2\beta\hbar}$, one shows that Eq. (3.2) reduces to 
$$i\hbar\,\tilde\xi_{BA}(t)=-{\beta\hbar\over 2}\,{d\tilde
C_{BA}(t)\over dt},\myeqno$$
which yields for the associated response function $\tilde\chi_{BA}(t,t')$ the expression
$$\tilde\chi_{BA}(t,t')=\beta\,\Theta(t-t')\,{\partial\tilde C_{BA}(t,t')\over\partial t'}.\myeqno$$

Eq. (3.5) constitutes the expression of the classical equilibrium fluctuation-dissipation theorem
in the time domain [3].
\bigskip
{\bf 3.3. Illustration : the harmonic oscillator}

One easily checks that Eq. (3.4) is actually verified by the functions $\tilde\xi_{xx}(t)$ as given
by Eq. (2.19) and $\tilde C_{xx}(t)$ as given by the classical limit of Eq. (2.20), namely
$$\tilde C_{xx}(t)={kT\over m\omega_0^2}\,\cos\omega_0t.\myeqno$$

\mysection{The zero temperature case}

{\bf 4.1. The zero-temperature fluctuation-dissipation theorem}

As the temperature decreases, the function $\log|\coth({\pi t/2\beta\hbar})|$ becomes more and
more spread around $t=0$.  At $T=0$, coming back to the formulation (2.16) of the FDT,
one gets
$$i\hbar\,\tilde\xi_{BA}(t)={1\over\pi}\,\tilde C_{BA}(t)*{\rm vp}\,{1\over t},\qquad\tilde
C_{BA}(t)=-{1\over\pi}\,i\hbar\,\tilde\xi_{BA}(t)*{\rm vp}\,{1\over t},\myeqno$$
that is
$$-i\hbar\,\tilde\xi_{BA}(t)={1\over\pi}\,{\rm vp}\int_{-\infty}^\infty\tilde
C_{BA}(t')\,{1\over t'-t}\,dt',\myeqno$$
and 
$$\tilde C_{BA}(t)=-{1\over\pi}\,{\rm vp}\int_{-\infty}^\infty(-i\hbar\,\xi_{BA}(t'))\,{1\over
t'-t}\,dt'.\myeqno$$

Interestingly enough, these Hilbert transformation relations, which constitute the formulation of
the fluctuation-dissipation theorem in the time domain at zero temperature, are formally similar to
the usual Kramers-Kronig relations between the real and imaginary parts of the generalized
susceptibility, except for the evident fact that they hold in the time domain and not in the
frequency domain. Otherwise stated, at $T=0$, the quantities $\tilde C_{BA}(t)$ and
$-i\hbar\,\tilde\xi_{BA}(t)$ must constitute respectively the real and imaginary parts of an
analytic signal $\tilde Z_{BA}(t)$ with only positive frequency Fourier components [9]-[10].
\bigskip
{\bf 4.2. The zero-temperature analytic signal}

Following these lines, consider the signal $\tilde Z_{BA}(t)=\langle B(t)A\rangle$. One has
$$\tilde Z_{BA}(t)=\tilde C_{BA}(t)+\hbar\,\tilde\xi_{BA}(t),\myeqno$$
which displays the fact that $\tilde Z_{BA}(t)$ has for real part $\tilde C_{BA}(t)$ and for
imaginary part $-i\hbar\,\tilde\xi_{BA}(t)$. By Fourier transformation, one gets
$$Z_{BA}(\omega)=C_{BA}(\omega)+\hbar\,\xi_{BA}(\omega).\myeqno$$

At $T=0$, the fluctuation-dissipation relations (2.15) reduce to
$$\xi_{BA}(\omega)={1\over\hbar}\,{\rm sgn}(\omega)\,C_{BA}(\omega),\myeqno$$
with
$${\rm sgn}(\omega)=\cases{+1,&$\omega>0$,\cr
-1,&$\omega<0$,\cr}\myeqno$$
so that one gets, as expected,
$$Z_{BA}(\omega)=\cases{2\,C_{BA}(\omega),&$\omega>0$,\cr
0,&$\omega<0$.\cr}\myeqno$$
Actually, at $T=0$, the function $\tilde Z_{BA}(t)=\langle B(t)A\rangle$ has only positive
frequency components and thus possesses the characteristics of an analytic signal [9]-[10]. This
implies that the integral definition $\tilde
Z_{BA}(\tau)=\int_0^\infty Z_{BA}(\omega)e^{-i\omega\tau}\,{d\omega/2\pi}$ can then be extended in
the whole lower half of the complex $\tau$-plane\footnote{$^3$}{This is in accordance with the
above noted fact that at finite temperature the prolongation to the region $-\beta\leq\IM\tau\leq
0$ of the complex $\tau$-plane of the correlation function $\langle B(\tau)A\rangle$ is analytic.}
({\sl i.e.\/} $\IM\tau\leq 0$). 

In the same way, consider the function $\tilde Y_{BA}(t)=\langle AB(t)\rangle$. One has:
$$\tilde Y_{BA}(t)=\tilde C_{BA}(t)-\hbar\,\xi_{BA}(t).\myeqno$$
At $T=0$, making use of the fluctuation-dissipation relations (2.15), one gets
$$Y_{BA}(\omega)=\cases{0,&$\omega>0$,\cr
2\,C_{BA}(\omega),&$\omega<0$.\cr}\myeqno$$
Thus, the function $\tilde Y_{BA}(t)=\langle AB(t)\rangle$ possesses only negative frequency
components.  The integral definition $\tilde Y_{BA}(\tau)=\int_0^\infty Y_{BA}(\omega)e^{-i\omega
\tau}{d\omega/2\pi}$ can be extended in the whole upper half of the complex $\tau$-plane ({\sl
i.e.\/} $\IM\tau\geq 0$).
\bigskip
{\bf 4.3. Other representations of the analytic signal}

Let us here focus on the analytic signal $\tilde Z_{BA}(t)$ of positive frequency components 
(similar considerations can be made for $\tilde Y_{BA}(t)$).

For $\IM\tau\leq 0$, one can write, taking
advantage of Eqs. (4.6) and (4.8),
$$\tilde Z_{BA}(\tau)={1\over 2\pi}\int_0^\infty
d\omega\,2\hbar\,\xi_{BA}(\omega)\,e^{-i\omega\tau}.\myeqno$$
This yields the following representation of $\tilde Z_{BA}(\tau)$ for $\IM\tau\leq 0$ in terms of
the dissipative part $\tilde\xi_{BA}(t)$ of the response function:
$$\tilde Z_{BA}(\tau)={1\over\pi}\int_{-\infty}^\infty
dt'\,(-i\hbar\,\tilde\xi_{BA}(t'))\,{1\over\tau-t'}.\myeqno$$
Note that, when $\IM\tau=0$, the integral in Eq. (4.12) must be understood as a principal value.

Eq. (4.12) can in turn be used as a definition of $\tilde Z_{BA}(\tau)$ in the upper half of the
complex $\tau$-plane ({\sl i.e.\/} $\IM\,\tau>0$), where the integral definition $\tilde
Z_{BA}(\tau)=\int_0^\infty Z_{BA}(\omega)e^{-i\omega\tau}{d\omega/2\pi}$ cannot be used. It
verifies the property
$$Z_{BA}^*(\tau)=Z_{BA}(\tau^*).\myeqno$$
\bigskip
{\bf 4.4. Illustration: the harmonic oscillator}

Let us consider once again the harmonic oscillator of mass $m$, angular frequency
$\omega_0$ and displacement $x$, in thermal equilibrium at $T=0$. 

At $T=0$, the equilibrium correlation functions $\tilde Z_{xx}(t)=\langle
x(t)x\rangle$ and $\tilde Y_{xx}(t)=\langle xx(t)\rangle$ are given by
$$\tilde Z_{xx}(t)={\hbar\over 2m\omega_0}\,e^{-i\omega_0t}\myeqno$$
and
$$\tilde Y_{xx}(t)={\hbar\over 2m\omega_0}\,e^{i\omega_0t}.\myeqno$$
Thus $\tilde Z_{xx}(t)$ is a monochromatic signal of angular frequency $\omega_0$ and of
Fourier spectrum
$$Z_{xx}(\omega)={\hbar\over m\omega_0}\,\pi\,\delta(\omega-\omega_0),\myeqno,$$
while, similarly, $\tilde Y_{xx}(t)$ is a monochromatic signal of angular frequency $-\omega_0$
and of Fourier spectrum
$$Y_{xx}(\omega)={\hbar\over m\omega_0}\,\pi\,\delta(\omega+\omega_0).\myeqno$$

The Hilbert transformation relations (4.1) between $\tilde C_{xx}(t)$ and 
$-i\hbar\,\tilde\xi_{xx}(t)$ read:
$$\sin\omega_0t={1\over\pi}\,\cos\omega_0t*{\rm
vp}\,{1\over t},\qquad\cos\omega_0t=-{1\over\pi}\,\sin\omega_0t*{\rm
vp}\,{1\over t}.\myeqno$$

The representation (4.12) of $\tilde Z_{xx}(\tau)$ in the complex $\tau$-plane reads:
$$\tilde Z_{xx}(\tau)=\cases{\display{\hbar\over 2m\omega_0}\,e^{i\omega_0\tau},&$\IM\tau>0$,\cr
\display{\hbar\over 2m\omega_0}\,e^{-i\omega_0\tau},&$\IM\tau\leq 0$.\cr}\myeqno$$

\mysection{Discussion and conclusion}

Clearly, the time domain formulation of the equilibrium fluctuation-dissipation theorem is
completely equivalent to the widely used frequency form of the theorem. In the classical limit, 
the time domain formulation establishes a proportionality
relation between the dissipative part of the response function and the derivative of the equilibrium
correlation function. At zero temperature, it takes the form of Hilbert transformation relations
between the dissipative part of the response function and the symmetrized equilibrium correlation
function. 

The time domain formulation is of considerable help in discussing out-of-equilibrium phenomena. A
good illustration of that can be found in the discussion of aging effects. For instance, in a recent
paper [8], we have studied these effects as displayed by the correlation function of the
displacement $x(t)-x(t_0)$ of a free quantum Brownian particle with respect to its position at a
given time $t_0$. Indeed, since diffusion is going on, the variable $x(t)-x(t_0)$ never attains
equilibrium. For any times $t$ and $t'$ such that $t_0\leq t\leq t'$, the displacement correlation
function $C_{xx}(t,t';t_0)$ depends both on the time difference $t-t'$ and on the waiting time or
age $t_w=t'-t_0$. Since the particle displacement cannot be viewed as corresponding to a stationary
stochastic process, Fourier analysis and the Wiener-Khintchine theorem cannot be used in order to
compute $C_{xx}(t,t';t_0)$. This quantity must thus be obtained through a double time integration
of the velocity correlation function $C_{vv}(t_1,t_2)$ (which only depends on the time difference
$t_1-t_2$ since the Brownian particle velocity thermalizes and does not age). 

In this situation, for any temperature of the thermal bath, one can write a modified FDT relating
the displacement response function $\chi_{xx}$ to the partial derivative ${\partial
C_{xx}(t,t';t_0)/\partial t'}$, this latter quantity taking into account
even those fluctuations of the particle displacement which take place during the waiting time (the
FDT being valid with no modifications only when $t_w=0$, {\sl i.e.\/} when one wants to relate
$\chi_{xx}$ and ${\partial C_{xx}(t,t';t_0)/\partial t'}\bigl|_{t_0=t'}$). This is rendered possible through the introduction of an
associated effective inverse temperature
$\beta_{\rm eff.}$ (or, equivalently, of a violation factor $X$) depending on both time arguments
$t-t'$ and $t_w$ [8]. 

The above considerations, which solely rely on the consideration of time-dependent quantities, 
can be extended to other out-of-equilibrium dynamic variables of dissipative systems, classical or
quantal [5]-[6].
\vfill
\break
\noindent
{\smalltitle Appendix A : some useful Fourier transforms and convolution relations}
\bigskip

{\bf A.1. Fourier transform of ${\bf tanh}\,{\bm{\beta\hbar\omega/2}}$}

Let us set
$$I_1=\int_0^\infty d\omega\,\sin\omega t\,\tanh{\beta\hbar\omega\over 2}.\eqno(A.1)$$ 
Using the expansion 
$$\tanh{\pi x\over 2}={4x\over\pi}\sum_{k=1}^\infty{1\over(2k-1)^2+x^2},\eqno(A.2)$$
with $\omega=({\pi/\beta\hbar})x$, one gets
$$I_1={2\pi\over\beta\hbar}\sum_{k=1}^\infty e^{-{(2k-1)\pi t/\beta\hbar}},\qquad
t>0,\eqno(A.3)$$ 
which yields the final result, valid whatever the sign of $t$:
$$I_1=\cases{\display{\pi\over\beta\hbar}\,{1\over\sinh{\pi t\over
\beta\hbar}},&$t\neq0$,\cr
0,&$t=0$,\cr}\eqno(A.4)$$
One thus has:
$${1\over 2\pi}\int_{-\infty}^\infty d\omega\,e^{-i\omega t}\,\tanh{\beta\hbar\omega\over
2}=-{i\over\pi}\,I_1.\eqno(A.5)$$
\bigskip

{\bf A.2. Fourier transform of ${\bf coth}\,{\bm{\beta\hbar\omega/2}}$}

Similarly, let us set
$$I_2=\int_0^\infty d\omega\,\sin\omega t\,\coth{\beta\hbar\omega\over 2}.\eqno(A.6)$$
Using the expansion
$$\coth{\pi x\over 2}={2\over\pi x}+
{4x\over\pi}\sum_{k=1}^\infty{1\over (2k)^2+x^2},\eqno(A.7)$$
with $\omega$ defined as above, one gets
$$I_2={\pi\over\beta\hbar}+{2\pi\over\beta\hbar}\sum_{k=1}^\infty
e^{-2k\pi t/\beta\hbar},\qquad t>0,\eqno(A.8)$$
which yields the final result, valid whatever the sign of $t$:
$$I_2=\cases{\display{\pi\over\beta\hbar}\coth{\pi t\over\beta\hbar}
,&$t\neq0$,\cr
0,&$t=0$,\cr}\eqno(A.9)$$
One thus has:
$${1\over 2\pi}\int_{-\infty}^\infty d\omega\,e^{-i\omega t}\,\coth{\beta\hbar\omega\over
2}=-{i\over\pi}\,I_2.\eqno(A.10)$$
\bigskip
{\bf A.3.}

From the relation
$$\tanh{\beta\hbar\omega\over 2}\,\coth{\beta\hbar\omega\over 2}=1,\eqno(A.11)$$
one deduces, by Fourier transformation and use of the above Fourier formulas (Eqs. ($A$.5) and
($A$.10)), the convolution relation
$${\pi\over\beta\hbar}\,{\rm vp}\,{1\over\sinh{\pi t\over\beta\hbar}}*{\pi\over\beta\hbar}\,{\rm
vp}\,\coth{\pi t\over\beta\hbar}=-\pi^2\,\delta(t).\eqno(A.12)$$

At $T=0$, Eq. ($A$.12) reduces to
$${\rm vp}\,{1\over t}*{\rm vp}\,{1\over t}=-\pi^2\,\delta(t).\eqno(A.13)$$

{\bf A.4. Some useful relations}

One has

$$\log|\coth\,ax|\sim_{|a|\to\infty}{\pi^2\over 4|a|}\,\delta(x)\eqno(A.14)$$
and
$${a\over\sinh ax}\sim_{|a|\to\infty}-{\pi^2\over 2|a|}\,\delta'(x).\eqno(A.15)$$

Formula ($A$.14) can be demonstrated in a standard fashion by considering $\log|\coth ax|$ as a
distribution. Indeed, for any well-behaved function $\phi(x)$, the integral
$\int_{-\infty}^\infty\log|\coth ax|\,\phi(x)\,dx$ tends towards
$({\pi^2/4|a|})\,\phi(0)$ in the limit $|a|\to\infty$.\break

Formulas ($A$.14) and ($A$.15) can be used to study the classical limit of the convolution 
relation ($A$.12). Indeed, in the classical limit, using Eq. ($A$.15) with
$a={\pi/2\beta\hbar}$, the l.h.s. of Eq. ($A$.12) is seen to reduce to
$-({\pi^2/2})\,\delta'(t)*{\rm sgn}(t)$, that is, to $-\pi^2\delta(t)$, as it should.
\vfill
\break
\noindent
{\smalltitle Appendix B : the harmonic oscillator case}
\bigskip
{\bf B.1. The first fluctuation-dissipation relation}

The first fluctuation-dissipation relation (2.21) expresses for the harmonic oscillator
the displacement response function in terms of the displacement correlation function. The
convolution product in the r.h.s. can be easily calculated -- and Eq.~(2.22) easily checked -- by
making use of the expansion
$${1\over\sinh{\pi t\over\beta\hbar}}=2\sum_{k=1}^\infty e^{-(2k-1)\pi t/\beta\hbar},\qquad
t>0,\eqno(B.1)$$
and computing separately each convolution product of the series.

In spite of their oscillating character, it is not necessary here to treat the oscillator
displacement response and correlation functions as distributions. Correspondingly, due to the fact
that the function
${1/\sinh}({\pi t/\beta\hbar})$ involved in the computation of the integral $I$  (Eq. (2.2)) tends
towards zero for large values of its argument, the contour integration on the contour $\Gamma$
described in Section 2.1 can be carried out without problems, even in the limit $R\to\infty$.
\bigskip
{\bf B.2. The second fluctuation-dissipation relation}

The second fluctuation-dissipation relation (2.22) expresses for the harmonic oscillator
the displacement correlation function in terms of the displacement response function. In order to
compute the convolution product in the r.h.s., it is convenient to write 
$$\coth{\pi t\over\beta\hbar}={\rm sgn}(t)+\bigl(\coth{\pi t\over\beta\hbar}-{\rm
sgn}(t)\bigr),\eqno(B.2)$$
with
$${\rm sgn}(t)=\cases{+1,&$t>0$,\cr
-1,&$t<0$,\cr}\eqno(B.3)$$
and to make use of the expansion
$$\coth{\pi t\over\beta\hbar}-{\rm sgn}(t)=\cases{\display 2\sum_{k=1}^\infty e^{-2k\pi
t/\beta\hbar},&$t>0$,\cr 
\display -2\sum_{k=1}^\infty e^{2k\pi t/\beta\hbar},&$t<0$.\cr}\eqno(B.4)$$
The first contribution to the r.h.s., namely the convolution product 
$$-{1\over\pi}\sin\omega_0
t*{\pi\over\beta\hbar}\,{\rm sgn}(t),\eqno(B.5)$$
only makes sense as a relation between distributions\footnote{$^4$}{This amounts to say that the
oscillating functions $\sin\omega_0 t$ and $\cos\omega_0 t$ have to be defined through the usual
limiting procedures, namely
$$\sin\omega_0 t=\lim_{\epsilon\to 0^+}\sin\omega_0 t\,e^{-\epsilon|t|},\qquad\cos\omega_0
t=\lim_{\epsilon\to 0^+}\cos\omega_0 t\,e^{-\epsilon|t|}.$$}. One has:
$$-{1\over\pi}\sin\omega_0 t*{\rm sgn}(t)={2\over\beta\hbar\omega_0}\,\cos\omega_0t.\eqno(B.6)$$
As for the second contribution to the r.h.s., namely the convolution product
$$-{1\over\pi}\sin\omega_0 t*{\pi\over\beta\hbar}\,\bigl(\coth{\pi t\over\beta\hbar}-{\rm
sgn}(t)\bigr),\eqno(B.7)$$
it can be easily calculated by making use of the expansion ($B$.4) and computing separately each
convolution product of the series. One thus gets:
$$-{1\over\pi}\sin\omega_0 t*{\pi\over\beta\hbar}\bigl(\coth{\pi t\over\beta\hbar}-{\rm
sgn}(t)\bigr)=\bigl(\coth{\beta\hbar\omega_0\over
2}-{2\over\beta\hbar\omega_0}\bigr)\,\cos\omega_0 t.\eqno(B.8)$$
Gathering together results ($B$.6) and ($B$.8), one obtains Eq. (2.22). Note that, interestingly
enough, it is only in the computation of the first contribution to the r.h.s. of Eq. (2.22), that
is, of the convolution product ($B$.5), that the consideration of the oscillator displacement
response and correlation functions as distributions is required.

Coming back to the computation of the contour integral
$J$ (Eq. 2.9)), one writes, for finite $R$,
$$J=2i\pi\tilde C_{xx}(t-t')+J_a+J_b.\eqno(B.9)$$
In Eq. ($B$.9), $2i\pi\tilde C_{xx}(t-t')$ is the contribution of the two small semicircles
centered in $t'$ and $t'-i\hbar\beta$, and $J_a$  denotes the contribution to
$J$ of the two vertical segments of abscissas $R$ and $-R$ (Fig.~1). As for $J_b$, it is defined by
$$J_b={\rm vp}\int_{-R}^R dt''\bigl[\langle
x(t'')x(t)\rangle-\langle
x(t)x(t'')\rangle\bigr]\,{\pi\over\beta\hbar}\,\coth{\pi(t''-t')\over\beta\hbar}.\eqno(B.10)$$
Interestingly enough, while $J_a$ and $J_b$, considered separately, do not tend towards a limit when
$R\to\infty$ but display instead an oscillatory behaviour, the sum $J_a+J_b$ possesses a well
defined limit. Indeed one has, for $R$ finite but such that $R\gg t',\beta\hbar$,
$$J_a\simeq -{2i\pi\over m\omega_0^2\beta}\,\cos\omega_0R\,\cos\omega_0 t,\eqno(B.11)$$
and
$$\displaylines{J_b\simeq {2i\pi\over m\omega_0^2\beta}\,\cos\omega_0R\,\cos\omega_0 t-{2i\pi\over
m\omega_0^2\beta}\,\cos\omega_0(t-t')\hfill\cr
\hfill +i{\hbar\over m\omega_0}\,{\rm vp}\int_{-R}^R
\sin\omega_0(t-t'')\,{\pi\over\beta\hbar}\bigl(\coth{\pi(t''-t')\over\beta\hbar}-{\rm
sgn}(t''-t')\bigr).\quad(B.12)\cr}$$
When $R\to\infty$, the sum $J_a+J_b$ possesses a well-defined limit, namely
$$\displaylines{J_a+J_b=-{2i\pi\over m\omega_0^2\beta}\,\cos\omega_0(t-t')\hfill\cr
\hfill +i{\hbar\over m\omega_0}\,
{\rm vp}\int_{-\infty}^\infty
\sin\omega_0(t-t'')\,{\pi\over\beta\hbar}\bigl(\coth{\pi(t''-t')\over\beta\hbar}-{\rm
sgn}(t''-t')\bigr).\quad(B.13)\cr}$$
Using then formula ($B$.5), and the fact that $J=0$, together with the expression~
(2.20) for $\tilde C_{xx}(t)$, one checks again, as expected, the fluctuation-dissipation relation
(2.22).
\vfill
\break
\parindent=0pt
{\smalltitle Figure caption}
\bigskip
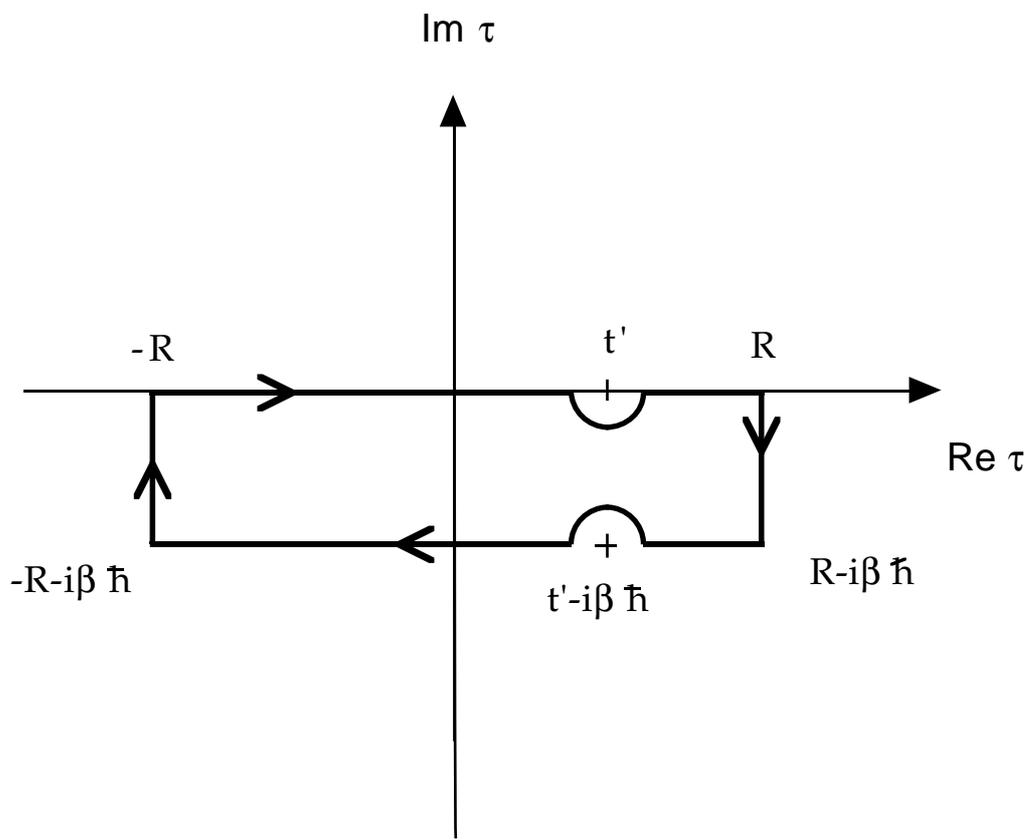
{\bf Fig.~1} 

Integration contour for the calculation of the integrals $I$ (Eq. (2.2)) and $J$\break (Eq. (2.9)).
\vfill
\break
\parindent=0pt
{\smalltitle References}
\bigskip
\baselineskip=12pt
\frenchspacing
1. H.B. Callen and T.A. Welton, Phys. Rev. {\bf 83}, 34 (1951).

2. H.B. Callen and R.F. Greene, Phys. Rev. {\bf 86}, 702 (1952).

3. R. Kubo, J. Phys. Soc. Japan {\bf 12}, 570 (1957).

4. R. Kubo, Rep. Prog. Phys. {\bf 29}, 255 (1966).

5. L.F. Cugliandolo, J. Kurchan and G. Parisi, J. Phys. {\bf 4}, 1641 (1994).

6. J.-P. Bouchaud, L.F. Cugliandolo, J. Kurchan and M. M\'ezard, in {\sl Spin-glasses and random
fields\/}, A.P. Young Ed. (World Scientific, 1997).

7. L.F. Cugliandolo and G. Lozano, Phys. Rev. Lett. {\bf 80}, 4979 (1998);\break Phys. Rev. B {\bf
59}, 915 (1999).

8. N. Pottier and A. Mauger, preprint cond-mat/9912028, to appear in Physica A.

9. J.W. Goodman, {\sl Statistical optics\/}, Wiley (1985).

10. L. Mandel and E. Wolf, {\sl Optical coherence and quantum optics\/}, Cambridge
University Press (1995).

11. S.W. Lovesey, {\it Condensed matter physics: dynamic correlations\/}, The Benjamin/Cummings
Publishing Company (1980).
\bye